\documentstyle{article}

\renewcommand{\d}{\rm d}
\newcommand{\card}{\rm card}
\newcommand{\Qu}{\rm Qu}
\newcommand{\ar}{\longrightarrow}
\newcommand{\w}{\omega}
\newcommand{\la}{\lambda}
\renewcommand{\a}{\alpha}
\begin{document}
\title{Quantum Computer Can Not Speed Up Iterated Applications
 of a Black Box}
\author{Yuri Ozhigov}
\maketitle

\ \ \ \ \ Address: \ Department of mathematics, Moscow state technological

\ \ \ \ \ University "Stankin", Vadkovsky per. 3a, 101472, Moscow, Russia

\ \ \ \ \ E-mail: \ y@oz.msk.ru

\section{Summary}

Let a classical algorithm be determined by sequential applications of a
black box performing one step of this algorithm. If we consider
this black box as an oracle which gives a value $f(a)$ for a query $a$,
we can compute $T$ sequential applications of $f$ on a classical computer 
relative to 
this oracle in time $T$. 

It is proved that if $T=O(2^{n/7} )$, where $n$ is the length of input, then
the result of $T$ sequential applications of $f$ can not be computed on quantum 
computer with oracle
for $f$ for all possible $f$ faster than in time $\Omega (T)$. This means that
there is no general method of quantum speeding up of classical algorithms 
provided in such a general method a classical algorithm is regarded as 
iterated applications of a given black box.

For an arbitrary time complexity $T$ a lower bound for the time of quantum 
simulation was found to be $\Omega (T^{1/2} )$.

\section{Introduction}

In the last years many investigators have amassed a convincing body
of evidence that a quantum device can be more powerful tool for computations
than a classical computer. This is because for the different problems 
there exist quantum algorithms which find a solution substantially faster
than any known (or even any possible) classical algorithm (look, for example,
at the works \cite{DJ},\cite{BB},\cite{Sh}).
The latest advance in quantum speeding up is the method of quantum
search proposed by L.Grover 
in the work \cite{Gr}. His algorithm takes  $O(\sqrt{N} )$
time when the classical search requires $\Omega (N)$ time. In some 
particular cases
(look in \cite{FG}) the time $O(\sqrt{N} )$ for a search can be even 
reduced. It would be natural to expect that some more general method of  
quantum speeding up can take place for all classical algorithms with time complexity
more than $O(n)$. 

One of the main general corollaries from
the classical theory of algorithms
is that if we know only a code of algorithm then in general case
the unique way to learn a result of computations is to run this 
algorithm on a given input. Therefore, given a code of algorithm, 
generally speaking we can only use it as a black box to perform sequentially
 all steps of computations and no other analysis can yield their result. 
Thus we can regard a computation $X_0 \ar X_1 \ar\ldots\ar X_T$
as iterated application of the same oracle $f$ which gives sequentially :
 $X_{s+1} =f(X_s ) ,\ s=0,1,\ldots ,T-1,\ T=T(n)>O(n)$.

In view of this we assume that a general method of quantum speeding up
of classical algorithms is a quantum query machine with oracle $f$ which 
yields the result $X_T$ of computations in time $\a (T)$, where 
$\a (T)/T \ar 0 \ (T\ar \infty )$. However, we shall see that such a method 
does not exist. This demonstrates a value of every partial result
about quantum speeding up because such results are all that can be done. 
(Though, the problem of quantum speeding up of all long computations
with time $>O( 2^{n/7} )$ still remains.)

Oracle quantum computers will be treated here within the framework of approach
proposed by C.Bennett, E.Bernstein, G.Brassard and U.Vazirani in the work
\cite{BBBV} . They considered a quantum Turing machine with oracle as a 
model 
of quantum computer. In this paper we use slightly different model of 
quantum computer with separated quantum and classical parts, but the 
results hold also for QTMs. We proceed with the exact definitions.

\section{Quantum computer with the separated quantum and classical 
parts}

Our quantum query machine consists of two parts: quantum and classical.
Let $\w^*$ denotes the set of all words in alphabet $\w$.

{\bf Quantum part.}

It consists of two infinite tapes: working and query, the finite set
$\cal U$ of unitary transformations which can be easily performed by
the physical devices, and infinite set $F=\bigcup\limits_{n=1}^\infty 
F_n$ of unitary transformations called an oracle for the
length preserving function
$f:\ \{ 0,1\}^* \ar\{ 0,1\}^*$, each $F_n$ acts on $2^{2n}$ dimensional 
Hilbert space spanned by $\{ 0,1\}^{2n}$ as follows: 
$F_n | \bar a,\bar b \rangle = | \bar a ,f(\bar a) \bigoplus \bar b \rangle$,
$\bar a ,\bar b \in\{ 0,1\}^n$, where $\bigoplus$ denotes the bitwise addition
modulo 2.

The cells of tapes are called qubits. Each qubit takes values from the 
complex 1-dimensional sphere of radius 1: $\{ z_0 {\bf 0} +z_1 {\bf 1} \ 
| \ z_1 ,z_2 \in {\tt C}, |z_0 |^2+|z_1 |^2
 =1\}$. Here $\bf 0$ and $\bf 1$ are referred as basic states of qubit 
and form the basis of ${\tt C}^2$. 

During all the time of computation the both tapes are limited each by two 
markers with fixed positions, so that on the working (query) tape only
qubits $v_1 ,v_2 ,\ldots ,v_T$ ($v_{T+1} ,v_{T+2} ,\ldots ,v_{T+2n}$)
are available in a computation with time complexity $T=T(n)$ 
on input of length $n$. Put $Q=\{ v_1 ,v_2 ,\ldots ,v_{T+2n} \}$.
A basic state of quantum part is a function of the form $e:\ Q\ar\{ 0,1\}$.
Such a state can be encoded as $|e(v_1 ) ,e(v_2 ) ,\ldots ,e(v_{T+2n}) 
\rangle$ and naturally identified with the corresponding word in alphabet
$\{ 0,1\}$. Let $K=2^{T+2n}$; $\ e_0 ,e_1 ,\ldots ,e_{K-1}$ be all basic 
states taken in some fixed order, $\cal H$ be $K$ dimensional Hilbert
space with orthonormal basis $e_0 ,e_1 ,\ldots ,e_{K-1}$. $\cal H$ can
be regarded as tensor product ${\cal H}_1 \bigotimes {\cal H}_2 \bigotimes
\ldots \bigotimes{\cal H}_{T+2n}$ of 2 dimensional spaces, where ${\cal H}_i$
is generated by all possible values of $v_i ,\ i=1,2,\ldots ,T+2n$.
A (pure) state of quantum part is such an element $x\in\cal H$ that $|x|=1$.

Time evolution of quantum part at hand is determined by two types of 
unitary transformations on its states: working and query.
Let a pair $G,U$ be somehow selected, where $G\subset\{ 1,2,\ldots ,T+2n\}$, 
$U\in\cal U$ is unitary transform on $2^{\card (G)}$ dimensional Hilbert space.

{\it Working transform} $W_{G,U}$ on $\cal H$ has the form $E\bigotimes U'$,
where $U'$ acts as $U$ on $\bigotimes\limits_{i\in G} {\cal H}_i$ in the basis
at hand, $E$ acts as identity on $\bigotimes\limits_{i\notin G} {\cal H}_i$.

{\it Query transform} $\Qu _f$ on $\cal H$ has the form $E\bigotimes F'_n$,
where $F'_n$ acts as $F_n$ on $\bigotimes\limits_{i=T+1}^{T+2n} {\cal H}_i$ and 
$E$ acts as identity on $\bigotimes\limits_{i=1}^T {\cal H}_i$.

{\it Observation} of the quantum part. If the quantum part is in state
$\chi =\sum\limits_{i=0}^{K-1} \la_i e_i ,$ an observation is a procedure which gives
the basic state $e_i$ with probability $|\la _i |^2$.

{\bf Classical part}. It consists of two classical tapes: working and query,
which cells are in one-to-one correspondence with the respective qubits of 
the quantum tapes and have boundary markers on the corresponding positions.
Every cell of classical tapes contains a letter from some finite
 alphabet $\w$.
Evolution of classical part is determined by the classical Turing machine $M$
with a few heads on both tapes and the set of integrated states of heads:
$\{ q_b ,q_w ,q_q ,q_o ,\ldots\}$. We denote by $h(C)$ the
integrated state of heads for a state $C$ of classical part.

Let $D$ be the set of all states of classical part. 

{\it Rule of correspondence } between quantum and classical parts has the form
$R:\ D\ar 2^{\{ 1,2,\ldots ,T+2n\} } \times\cal U$, where $\forall C\in D$
$R(C)=\langle G,U\rangle$, $U$ acts on $2^{\card (G)}$ dimensional Hilbert 
space so that $U$ depends only on $h(C)$, and 
the elements of $G$ are exactly the numbers of those cells on classical tape
which contain the special letter $a_0 \in\w$.  

A state of quantum computer at hand is a pair $S=\langle Q(S),C(S)\rangle$
where $Q(S)$ and $C(S)$ are the states of quantum and classical parts respectively.

{\it Computation} on quantum computer. It is a chain of transformations of the 
following form:
\begin{equation}
S_0 \ar S_1 \ar \ldots \ar S_T ,
\label{1}\end{equation}
where for every $i=0,1,\ldots ,T-1$ $C(S_i )\ar C(S_{i+1} )$ is transformation 
determined by Turing machine M, and the following properties are fulfilled:

if $h(C(S_i ))=q_w$ then $Q(S_{i+1} )=W_{R(C(S_i ))} (Q(S_i ))$,

if $h(C(S_i ))=q_q$ then $Q(S_{i+1} )=\Qu_f (Q(S_i ))$,

if $h(C(S_i ))=q_b$ then $i=0$, $Q(S_0 )=e_0 ,\ C(S_0 )$ is fixed initial state,
corresponding to input word $a\in \{ 0,1\}^n$,

if $h(C(S_i ))=q_o$ then $i=T$,

in other cases $Q(S_{i+1} )=Q(S_i )$.

We say that this quantum computer (QC) computes a function $F(a)$ with probability 
$p\geq 2/3$ and time complexity $T$ if for the computation (\ref{1}) on every input
$a$ the observation of $S_T$ and the following routine procedure fixed
beforehand give $F(a)$ with probability $p$. We always can reach any other value
of probability $p_0 >p$ if fulfill computations repeatedly on the same input
and take the prevailing result. This leads only to a linear slowdown of computation.

\section{The effect of changes in oracle on the result of quantum 
computation}

For a state $e_j =|s_1 , s_2 , \ldots , s_{T+2n} \rangle$ of the quantum part 
we denote
the word $s_{T+1} s_{T+2} \ldots s_{T+n}$ by $q(e_j )$.
The state $S$ of QC is called query if $h(C(S))=q_q$. Such a state is querying the 
oracle
on all the words $q(e_j )$ with some amplitudes.
 Put ${\cal K} =\{ 0,1,\ldots , K-1\}$. Let $\xi=
Q(S)=\sum\limits_{j\in\cal K} \la_j e_j$. 
Given a word $a\in\{ 0,1\}^n$ for a 
query state $S$ we define:
$$
\delta_a (\xi )=\sum\limits_{j:\ q(e_j )=a} |\la_j |^2 .
$$

It is the probability that a state $S$ is querying the oracle on the word $a$.
In particular, $\sum\limits_{a\in\{ 0,1\}^n } \delta_a (\xi ) =1$.

Each query state $S$ induces the metric on the set of all oracles if for 
length preserving functions $f,g$ we define a distance between them by 
$$
\d_S (f,g)= ({\sum\limits_{a:\ f(a)\neq g(a)} \delta_a (\xi ) } )^{1/2} .
$$

\newtheorem{Lemma}{Lemma}
\begin{Lemma} Let $\Qu _f ,\ \Qu_g$ be query transforms on quantum part of QC
corresponding to functions $f,g$; $S$ be a query state. Then
$$
|\Qu_f (S) -\Qu_g (S)|\leq 2\d_S (f,g) .
$$
\end{Lemma}

{\bf Proof}

Put ${\cal L} =\{ j\in{\cal K} \ |\ f(q(e_j ))\neq g(q(e_j ))\}$.
We have: $|\Qu_f (S) -\Qu_g (S) |\leq 2(\sum\limits_{j\in\cal L} (|\la_j |)^2 )^{1/2}
\leq2\d_S (f,g) .\ \Box$

Now we shall consider the classical part of computer as a part of working tape.
Then a state of computer will be a point in $K^2$ dimensional Hilbert space
${\cal H}_1$. We denote such states by $\xi ,\chi$ with indices. All transformations
of classical part can be fulfilled reversibly as it is shown by C.Bennett
in the work \cite{Be}. This results in that all transformations in computation
(\ref{1}) will be unitary transforms in ${\cal H}_1$. At last we can 
 join sequential steps: $S_i \ar S_{i+1} \ar\ldots\ar S_j$ where $S_i \ar S_{i+1}$,
$S_j \ar S_{j+1}$ are two nearest query transforms, in one step. So
the computation on our QC acquires the form

\begin{equation}
\chi_0 \ar \chi_1 \ar \ldots \ar \chi_t ,\label{2}
\end{equation}
where every passage is the query unitary transform and the following 
unitary transform $U_i$ which depends only on $i$: 
$\ \chi_i \stackrel{\Qu_f }{\ar} \chi '_i 
\stackrel{U_i}{\ar} \chi_{i+1}$. We shall denote $U_i (\Qu_f (\xi ))$ 
by $V_{i,f} (\xi )$, then $\chi_{i+1} =V_{i,f} (\chi_i ),\ i=0,1,\ldots ,t-1$.
Here $t$ is the number of query transforms (or evaluations of the function $f$)
in the computation at hand.
Put $\d_a (\xi )=\sqrt{\delta_a (\xi )}$.
\begin{Lemma}
If $\xi _0 \ar\xi_1 \ar\ldots\ar\xi_t$ is a computation with oracle for $f$, a 
function $g$ differs from $f$ only on one word $a\in\{ 0,1\}^n$ and 
$\xi  _0 \ar\xi '_1 \ar\ldots\ar\xi '_t$ is a computation on the same QC with a 
new oracle for $g$, then
$$
|\xi _t -\xi '_t |\leq 2\sum\limits_{i=0}^{t-1} \d_a (\xi_i ) .
$$
\end{Lemma}

{\bf Proof}

Induction on $t$. Basis is evident. Step. 
In view of that $V_{t-1,g}$ is unitary, Lemma 1 and 
inductive hypothesis,
we have
$$
\begin{array}{l}
|\xi_t -\xi '_t |=|V_{t-1,f} (\xi_{t-1} )-V_{t-1,g} (\xi ' _{t-1} )|\leq \\
|V_{t-1,f} (\xi _{t-1} )-V_{t-1,g} (\xi_{t-1} )|+|V_{t-1,g} (\xi_{t-1} )-V_{t-1,g} 
(\xi '_{t-1} )| \leq \\
2\d_a (\xi _{t-1} ) +|\xi _{t-1} -\xi '_{t-1} |=
2\d_a (\xi _{t-1} ) +
2\sum\limits_{i=0}^{t-2}\d_a (\xi_i )=
2\sum\limits_{i=0}^{t-1}\d_a 
(\xi_i ) . \ \Box
\end{array}
$$

\section{Main results}

For a length preserving function $f$ a result of its iteration $f^{\{ k\} }$
is defined by the induction on $k$: $f^{\{ 0\} }$ is identity mapping,
$f^{\{ k+1\} } (x)=f(f^{\{ k\} } (x))$.

\newtheorem{Theorem}{Theorem}
\begin{Theorem} There is no such QC with oracle for $f$ that for some functions
$t(n) ,T(n):$ $t(n)/T(n) \ar 0\ (n\ar\infty )$, $T(n)=O(2^{n/7} )$ 
and every $f$ QC computes
$f^{\{T(n)\} } (\bar 0 )$ applying only $t(n)$ evaluations of $f$.
\end{Theorem}

{\bf Proof}

Suppose that it is not true and some QC with oracle for $f$ computes 
$f^{\{T(n)\} } (\bar 0 )$
applying only $t(n)$ evaluations of $f$, 
where $t(n)/T(n) \ar 0\ (n\ar\infty )$, $T(n)=O(2^{n/7} )$, 
and obtain a contradiction.

Let $f:\ \{ 0,1\}^* \ar\{ 0,1\}^*$  be such length preserving bijection
that for every $n=1,2,\ldots $ the orbit of the word $\bar 0 =0^n$ contains all
words from $\{ 0,1\}^n$. Let an oracle for $f$ be taken for the computation 
of $f^{\{T\} } (\bar 0 )$ on our QC.
This computation has the form (\ref{2}) where $t/T\ar 0\ (n\ar\infty )$. 
Let $n$ be sufficiently large so that $2t<T$.

Now we shall define the lists of the form $\langle \xi_i ,f_i ,{\cal T}_i 
, x_i \rangle$ where $\xi_i$ is a state from ${\cal H}_1$, $|\xi_i |=1$,
$f_i :\ \{ 0,1\}^* \ar\{ 0,1\}^*$ is length preserving function,
$x_i \in {\cal T}_i \subseteq \{ 0,1\}^n$ by the following induction on $i$.

Basis: $i=0$. Put $\xi_0 =\chi_0$, $f_0 =f,\ x_0 =\bar 0$, ${\cal T}_0 =\{ 0,1\}^n$.

Step. Put 
$$
\begin{array}{l}
\xi_{i+1} =V_{i,f_i} (\xi_i ) , \\
{\cal T}_{i+1} ={\cal T}_i \cap R_i ,\ \ R_i=\{a\ |\ \delta_a (\xi_{i+1} )<
\frac{1}{T^\a } \} ,
\end{array}
$$
$f_{i+1}$ differs from $f_i$ at most on one word $x_i$ where we define
$x_{i+1} =f_{i+1} (x_i )$ such that for all $s=1,2,\ldots ,T$
$f_{i+1}^{\{ s\} } (x_i ) \in{\cal T} _{i+1}$.

Note that $2^n -\card (R_i ) <T^\a .$ Therefore we can chose $x_{i+1}$ such that 
$x_{i+1} =f^{\{ j\} } (0)$ where $j<(i+1)TT^\a .$ It is possible for every 
$i=1,2,\ldots ,t-1$ if $\a\leq 5$ and $n$ is sufficiently large,
because $T=O(2^{\frac{n}{7}} )$.

We introduce the following notations: $V_i =V_{i,f_t} ,\ V_i^* =V_{i,f_i}$.
Let the unitary operator $V^i$ be introduced by the following induction:
$V^0 (x)=V_0 (x), \ V^i (x) =V_i (V^{i-1} (x))$, and the 
unitary operator $\tilde V_i$ be defined
by $\tilde V_0 =V_0^* ,$ $\tilde V_i (x) =V_i^* (\tilde V_{i-1} (x))$.
Then $\xi_{i+1} =\tilde V_i (\xi_0 )$. 

Put $\xi '_0 =\xi_0 ,\ \ \xi '_{i+1} =V^i (\xi _0 ) ,\ \ \partial _i =|\xi_i
-\xi '_i | ,\ \Delta_i =|V_i^* (\xi_i )-V_i (\xi_i )|$. It follows from the
definition that $f_i$ differs from $f_t$ at most on the set $X_i =
\{ x_i ,x_{i+1} ,\ldots ,x_{t-1} \}$ where $\forall a\in X_i\ \ \ \ 
\delta_a (\xi_i ) <\frac{1}{T^\a}$. Consequently, applying Lemma 1 
we obtain

\begin{equation}
\Delta_i \leq \frac{2t^{1/2}}{T^{\a /2}} .
\label{3}
\end{equation}

\begin{Lemma}
$\ \ \ \ \partial_i \leq\sum\limits_{k<i} \Delta _k .$
\end{Lemma}

{\bf Proof}

Induction on $i$. Basis follows from the definitions. Step:
$$
\begin{array}{l}
\partial_{i+1} =|\tilde V_i (\xi_0 ) -V^i (\xi_0 )|=|V_i^* (\tilde V_{i-1}
(\xi_0 ))-V_i (V^{i-1} (\xi_0 ))|\leq \\
\leq |V_i^* (\xi_i )-V_i (\xi_i )|+|V_i (\xi_i )-V_i (\xi '_i )|=\Delta_i +\partial_i .
\end{array}
$$
Applying the inductive hypothesis we complete the proof. $\Box$

Thus in view of (3) Lemma 3 gives
\begin{equation}
\forall i=1,\ldots ,t\ \ \ \partial_i \leq\frac{2it^{1/2}}{T^{\a /2}} .
\label{4}
\end{equation}

It follows from the definition of the functions $f_i$ that $\forall i\leq t \ \ 
\delta_{x_t} (\xi _i )<\frac{1}{T^\a } .$ Taking into account inequality (\ref{4} ),
we conclude that for $x=x_t$

$
\d_x (\xi_i -\xi '_i )\leq\frac{2it^{1/2}}{T^{\a /2}},$
$\d_x (\xi _i )<\frac{1}{T^{ \a /2}},$
$\d_x (\xi _i ' )\leq\d_x (\xi_i -\xi '_i )+
\d_x (\xi _i ).$

Hence we have
\begin{equation}
\d_x (\xi_i ')\leq
\frac{3t^{3/2}}{T^{\a /2}}. \label{5}
\end{equation}

Now we can change the value of the function $f_t$ only on the word
$x_t$ and obtain a new function $\phi$ such that 
$\phi^{\{T\}} (\bar 0 ) \neq f_t^{\{ T\}} (\bar 0 )$. Therefore, if
$\xi_0 \ar\xi_1 '' \ar\ldots \ar\xi_t ''$ is the computation of
$\phi^{\{ T\}} (\bar 0 )$ on our QC with oracle for $\phi$, then we have

\begin{equation}
|\xi '_t -\xi ''_t |\geq 1 .
\label{6}
\end{equation}

On the other hand, Lemma 2 and inequality (\ref{5}) give
$$
|\xi '_t -\xi ''_t |<2\sum\limits_{i\leq t} \d_x (\xi_i ' )
\leq\frac{6t^{5/2}}{T^{\a /2}} <1
$$
for $\a \geq 5$ and sufficiently large $n$, which contradicts to (\ref{6}).
Theorem 1 is proved.

\medskip
If the time complexity of classical computation exceeds $O(2^{n/7} )$ we can only
establish a lower bound for the time of quantum simulation as 
$\Omega (T^{1/2} )$. 

\begin{Theorem}
For arbitrary function $T(n)$ there is no such QC with oracle for $f$ that
for some function $t(n):\ \ t^2 /T \ar 0\ (n\ar\infty )$ QC computes 
$f^{\{T\}} (\bar 0 )$ for every $f$ applying only $t$ evaluations of $f$.
\end{Theorem}

{\bf Proof}

Let $f$ be selected as above.
Put $f^k =f^{\{ k\}} (\bar 0 ) \ \ k=0,1,\ldots ,T$. Define the matrix 
$A=(a_{ij} )$
with the following elements: $a_{ij} =\delta_{f^j} (\chi_i ),\ \ i=0,1,\ldots ,t;
\ j=0,1,\ldots ,T$. We have for every $i=0,\ldots ,t\ \ \sum\limits_{j=0}^T a_{ij} 
\leq 1$, consequently $t\geq\sum\limits_{i=0}^t \sum\limits_{j=0}^T a_{ij} =
\sum\limits_{j=0}^T \sum\limits_{i=0}^t a_{ij}$ and there exists such 
$\tau \in\{ 0,1,\ldots ,T\}$ that $\sum\limits_{i=0}^t a_{i\tau} \leq\frac{t}{T}$.

Changing the value of $f$ only on the word $f^\tau$ we obtain a new function
$g$ where $g ^{\{T\}} (\bar 0 ) \neq f^{\{ T\}} (\bar 0 )$.
Let $\chi_0 \ar\chi '_1 \ar\ldots \ar\chi '_t$ be computation on QC with oracle
for $g$. Then we have 
\begin{equation}
|\chi_t -\chi '_t |\geq 1 .
\label{7}
\end{equation}
On the other hand Lemma 2 gives
$
|\chi_t -\chi '_t |\leq 2\sum\limits_{i=0}^t \sqrt{a_{i\tau}} \leq2\sqrt{t
\sum a_{i\tau}} \leq t/T^{1/2} <1$ for sufficiently large $n$, which contradicts to
(\ref{7}). Theorem 2 is proved.

Note that for $T=\Omega (2^n )$ the lower bound as $\Omega (T^{1/2} )$
for the time of quantum simulation follows immediately from
the lower bound for the time of quantum search established in the
work \cite{BBBV}.

\section{Acknowledgments}

I am grateful to Charles H. Bennett who clarified for me some details of the work 
\cite{BBBV}, to Peter Hoyer for his comments and useful criticism and to 
Lov K. Grover for his attention to my work.


\begin{thebibliography}{99}
\bibitem[BBBV]{BBBV} C.H.Bennett, E.Bernstein, G.Brassard, U.Vazirani,
{\it Strenths and Weaknesses of Quantum Computing}, To appear in SIAM Journal on Computing (lanl e-print quant-ph/9701001)
\bibitem[BB]{BB} A.Berthiaume, G.Brassard, {\it Oracle quantum computing}, Journal of modern optics, 41(12):2521-2535, 1994 
\bibitem[Be]{Be} C.H.Bennett, {\it Logical reversibility of 
computation}, IBM J. Res.Develop. 17, 525-532
\bibitem[DJ]{DJ}, D.Deutsch, R.Jozsa, {\it Rapid solution of problems by quantum computation},
Proc. Roy. Soc. Lond. A {\bf 439} 553-558
\bibitem[FG]{FG} E.Farhi, S.Gutmann, {\it Quantum Mechanical Square Root Speedup
in a Structured Search Problem}, lanl e-print, quant-ph/9711035
\bibitem[Gr]{Gr} L.K.Grover, {\it A fast quantum mechanical algorithm for database search}, Proceedings, STOC 1996, Philadelphia PA USA, pp 212-219
\bibitem[Sh]{Sh} P.W.Shor, {\it Polynomial-Time Algorithms for Prime Factorization and Discrete Logarithms on Quantum Computer},
lanl e-print, quant-ph/9508027 v2 (A preliminary version in Proceedings of the 35th Annual Symposium on Foundations of Computer Science, Santa Fe, NM, Nov. 20-22, 1994, IEEE Computer Society Press, pp 124-134)
\end{thebibliography}
\end{document}